\documentclass[journal,transmag]{IEEEtran}
\usepackage{graphicx}   
\usepackage{hyperref}       
\usepackage{romannum}
\usepackage{amsmath}
\usepackage{mathtools}
\usepackage{algorithm} 
\usepackage{algpseudocode} 
\usepackage{amsfonts}
\usepackage{cite}
\usepackage{subcaption}
\usepackage[dvipsnames]{xcolor}
\usepackage{colortbl}
\usepackage{bm}
\usepackage{booktabs}

\newtheorem{thm}{Theorem}
\newtheorem{defn}{Definition}
\newtheorem{lem}{Lemma}
\newtheorem{assum}{Assumption}
\newtheorem{rem}{Remark}

\def\BibTeX{{\rm B\kern-.05em{\sc i\kern-.025em b}\kern-.08em
    T\kern-.1667em\lower.7ex\hbox{E}\kern-.125emX}}
\begin{document}
\title{Self-triggered Min-max DMPC for Asynchronous Multi-agent Systems with Communication Delays}
\author{Henglai Wei, 
Kunwu Zhang, 
Yang Shi 
\thanks{This work was supported by the Natural Sciences and Engineering Research Council of Canada (NSERC). (Corresponding author: Yang Shi.)

H. Wei, K. Zhang, Y. Shi are with the Department of Mechanical Engineering and the Institute for the Integrated Energy Systems (IESVic), University of Victoria, Victoria, BC, V8W 3P6, Canada (e-mail: henglaiwei@uvic.ca; kunwu@uvic.ca; yshi@uvic.ca).}}

\markboth{IEEE TRANSACTIONS ON INDUSTRIAL INFORMATICS,~Vol.~x, No.~x, x~2021}%
{Shell \MakeLowercase{\textit{et al.}}: Bare Demo of IEEEtran for IEEE Journals} 
\maketitle

\begin{abstract} 
This paper studies the formation stabilization problem of asynchronous nonlinear multi-agent systems (MAS) subject to parametric uncertainties, external disturbances and bounded time-varying communication delays. A self-triggered min-max distributed model predictive control (DMPC) approach is proposed to address this problem. At triggering instants, each agent solves a local min-max optimization problem based on local system states and predicted states of neighbors, determines its next triggering instant and broadcasts its predicted state trajectory to the neighbors. As a result, the communication load is greatly alleviated while retaining robustness and comparable control performance compared to periodic DMPC algorithms. In order to handle time-varying delays, a novel consistency constraint is incorporated into each local optimization problem to restrict the deviation between the newest predicted states and previously broadcast predicted states. Consequently, each agent can utilize previously predicted states of its neighbors to achieve cooperation in the presence of the asynchronous communication and time-varying delays. The proposed algorithm's recursive feasibility and MAS's closed-loop stability at triggering instants are proven. Finally, numerical simulations are conducted to verify the theoretical results.
\end{abstract}

\IEEEpeerreviewmaketitle
\section{Introduction}
Recently, the research on multi-agent systems (MAS) has attracted considerable attention due to its wide range of industrial applications, such as intelligent transportation systems \cite{liu2018distributed,guo2021systematic}, energy systems \cite{li2015distributed,rahman2017multi} and multi-robot systems \cite{feng2016robust}. However, these MAS may suffer from several challenging issues, such as the heavy communication burden, time-varying communication delays, and uncertainties. Hence, it is desirable to develop a robust control algorithm to simultaneously handle these issues of the MAS subject to state and input constraints.

For MAS, the periodic implementation of distributed control algorithms may lead to an undesirable communication burden and induce a nontrivial energy consumption. Alternatively, the \emph{event-based} distributed control method is a promising solution to reduce the communication load of MAS since the control inputs are only updated and transmitted to actuators at triggering time instants \cite{dimarogonas2011distributed,wang2020adaptive,wang2019distributed,tian2020memory}. In contrast with these methods, event-based distributed MPC (DMPC) has the advantage of systematically handling the constraints of MAS, reducing the communication burden, while optimizing the control performance. Based on different triggering conditions, the triggered DMPC in existing literature can be mainly classified into two categories: event-triggered \cite{gross2015cooperative,zou2019event,Kolarijani2020adencentralized}, and self-triggered DMPC \cite{mi2019self,zhan2018distributed}. The authors in \cite{zhan2018distributed} present a self-triggered DMPC algorithm for the consensus problem of synchronous linear MAS. However, these communication-efficient DMPC schemes result in the asynchronous information communication among agents, which may degrade the control performance (or even destroy the closed-loop stability) due to the inaccurate information of neighbors \cite{fang2005information}. The authors in \cite{mi2019self} propose a co-design self-triggered DMPC method for the linear MAS with asynchronous communication. To be specific, the previously broadcast information is used to estimate current optimal states; however, the estimation error induced by the asynchronous communication is not explicitly handled. Some promising results along the research line of the coordination of asynchronous MAS can be found in, for example, \cite{zou2019event,incremona2017asynchronous}. On the other hand, the delay-free transmission among MAS, assumed in the literature mentioned above, is impractical. Understandably, the time-varying communication delays may prevent agents from achieving the control objective. This issue has been studied in \cite{li2017robust,el2015distributed}. We note that most of these works only concentrate on one of these issues while ignoring others. Simultaneously addressing the aforementioned issues of MAS remains difficult, especially for nonlinear MAS with state and control constraints.

Another challenge is how to guarantee the robustness of the controlled nonlinear MAS when both additive disturbances and parametric uncertainties are presented. Min-max MPC has proven to be an effective method to address this problem \cite{limon2006input,lazar2008input}, where these two types of uncertainties can be tackled by minimizing the cost function related to the worst-case realizations of the uncertainties. To simultaneously overcome the above-mentioned challenges, we propose a self-triggered min-max DMPC for the nonlinear MAS. 

The main contribution of this work is threefold.
\begin{itemize}
\item A self-triggered min-max DMPC method is proposed for the asynchronous nonlinear MAS subject to {parametric uncertainties,} external disturbances and bounded time-varying delays. The proposed method can considerably reduce the communication burden and the frequency of solving the optimization problems while attaining the control performance comparable to the periodic DMPC.
\item A new consistency constraint that restricts the deviation between the current predicted and previously broadcast states is developed and incorporated into the local optimization problem. Consequently, agents achieve cooperation in the presence of time-varying delays.
\item The recursive feasibility of the algorithm is analyzed and the sufficient condition on guaranteeing the feasibility is provided. The closed-loop MAS are proven to be input-to-state practically stable (ISpS) at triggering instants.
\end{itemize}

{The remainder of this paper} is organized as follows: Section~\ref{whl2-sec:2} describes the formation stabilization control problem of nonlinear MAS. In Section~\ref{whl2-sec:3}, the optimization problem and the self-triggered scheduler are presented. Section~\ref{whl2-sec:4} provides the theoretical analysis of the recursive feasibility and the closed-loop stability. The simulation study is conducted in Section~\ref{whl2-sec:5}. Finally, the conclusion is given in Section~\ref{whl2-sec:6}.

The symbols $\mathbb{R}$, $\mathbb{R}_{\geq 0}$, $\mathbb{I}$ and $\mathbb{I}_{[m,n]}$ denote the sets of real numbers, nonnegative real numbers, the nonnegative integers and integers in $[m,n],m<n$, respectively. For ${x}\in \mathbb{R}^n$, $\|{x}\|$ denotes the Euclidean norm, $\|{x}\|_{{P}}$ denotes the weighted norm $\sqrt{{x}^\text{T}P{x}}$, where {the matrix} $P$ is positive definite. Given two sets $\mathcal{X},\mathcal{Y}\subseteq \mathbb{R}^n$, the set operation $\mathcal{X}\backslash \mathcal{Y}$ is defined by $\mathcal{X}\backslash \mathcal{Y}:=\{{x}\mid {x}\in\mathcal{X},{x}\notin \mathcal{Y}\}$. $\bar{\lambda}({P})$ and $\underline{\lambda}(P)$ denote the largest and smallest eigenvalues of the matrix $P$, respectively. ${x}(t)$ denotes the state ${x}$ at time $t$ and ${x}(s|t)$ denotes the predicted state at future time $t+s$ determined at time $t$, $s\in\mathbb{I}$. 

\section{Preliminaries and Problem Formulation}
\label{whl2-sec:2}
\subsection{Preliminaries}
Consider a discrete-time perturbed nonlinear system
\begin{equation}\label{whl2-eq:1}
{x}^+=g({x},{d}),
\end{equation}
where ${x}\in\mathbb{R}^n$ is the state, ${d}=[{w},{v}]\in\mathbb{D}\subset\mathbb{R}^d$ is the uncertainty, ${x}^+$ is the successor state and $g:\mathbb{R}^n\times\mathbb{R}^d\to \mathbb{R}^n$. $\mathbb{D}$ is a compact set and contains the origin in its interior. ${w}$ is the external disturbance and ${v}$ is the parametric uncertainty. 

Before formulating the control problem, we present the following definitions and lemma. 
\begin{defn}\cite{limon2006input}\label{whl2-def:1}
A set $\Omega\subset\mathbb{R}^n$ is {robustly positive} invariant (RPI) for the uncertain system ${x}^+=g({x},{d})$, if $g({x},{d})\in \Omega$, for all $ {x}\in \Omega$ and {all ${d}\in \mathbb{D}$}. 
\end{defn}

\begin{defn}\cite{limon2006input}\label{whl2-def:2}
For the system ${x}^+=g({x},{u},{d})$, a set $\Omega\subset\mathbb{R}^n$ is robust control invariant (RCI), if for all ${x}\in \Omega$, there exists an admissible control input ${u}$, such that $g({x},{u},{d})\in \Omega$, for all ${d}\in \mathbb{D}$. The $\ell$-step robustly stabilizable set $\mathbb{X}^{\ell}(\Omega)$, i.e., $\forall {x}\in \mathbb{X}^{\ell}(\Omega)$, the system can be robustly steered into $\Omega$ in $\ell$ steps, is denoted as $\mathbb{X}^{\ell}(\Omega):=\{{x}\in \mathbb{X} \mid \exists \, {u}\in\mathbb{U}$ such that $g({x},{u},\mathbb{D})\subseteq \mathbb{X}^{\ell-1}(\Omega)\}$, where $g({x},{u},\mathbb{D}):=\{g({x},{u},{d})\mid {d}\in \mathbb{D}\}$. In addition, $\mathbb{X}^0(\Omega)= \Omega$.
\end{defn}

\begin{lem}\cite{limon2006input}\label{whl2-lem:1}
For the system in \eqref{whl2-eq:1} with an RPI set $\Pi$, a function $V(\cdot):\mathbb{R}^n\to \mathbb{R}_{\geq 0}$ is called an ISpS Lyapunov function, if it satisfies
\begin{equation*}
\alpha_1(\|{x}\|)\leq V({x})\leq\alpha_2(\|{x}\|)+c_{1},
\end{equation*}
\begin{equation*}
V({x}^+)-V({x})\leq-\alpha_3(\|{x}\|)+\gamma(\|{w}\|)+c_{2},
\end{equation*}
for all ${x}\in \Pi$, ${d}\in \mathbb{D}$, where constants $c_{1},c_{2}\in\mathbb{R}_{\geq 0}$, $\alpha_1(\cdot)$, $\alpha_2(\cdot)$, $\alpha_3(\cdot)$ are $\mathcal{K}_\infty$ functions, and $\gamma(\cdot)$ is $\mathcal{K}$ function. If the system admits an ISpS Lyapunov function, then it is ISpS. 
\end{lem}

\begin{figure*}[!ht]
\centering
\begin{subfigure}[t]{0.32\textwidth}
\centering
\includegraphics[height=1.2in]{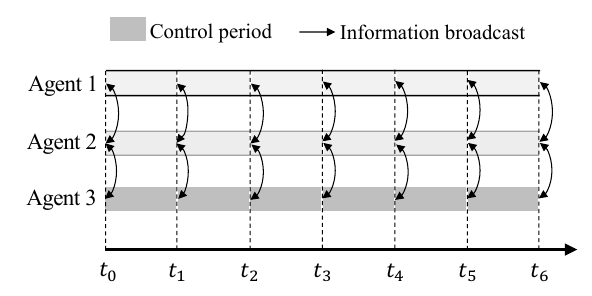}
\caption{Periodic and synchronous \cite{dunbar2007distributed}}
\end{subfigure}%
~ 
\begin{subfigure}[t]{0.32\textwidth}
\centering
\includegraphics[height=1.2in]{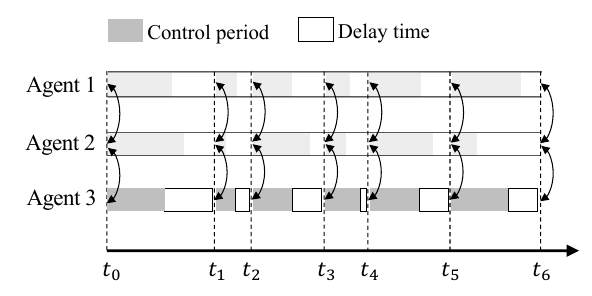}
\caption{Aperiodic and synchronous}
\end{subfigure}
\begin{subfigure}[t]{0.32\textwidth}
\centering
\includegraphics[height=1.2in]{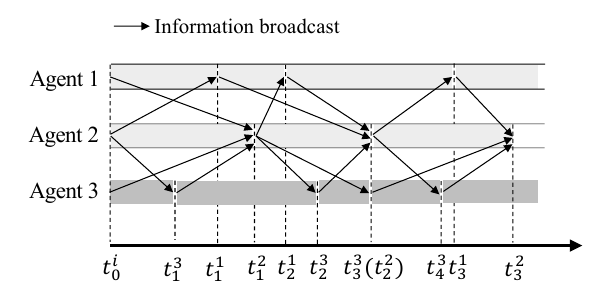}
\caption{Aperiodic and asynchronous}
\end{subfigure}
\caption{An example of MAS which consists of three agents: $\mathcal{N}_1=\{2\}, \mathcal{N}_2=\{1,3\},\mathcal{N}_3=\{2\}$. Periodic versus aperiodic and synchronous versus asynchronous DMPC methods are illustrated. For the synchronous DMPC method, all agents simultaneously update the states at time $t_k$, $k\in\mathbb{I}$. In (b), the update time $t_k$, $k\in\mathbb{I}$ is aperiodically synchronized based on the maximum communication delay. In (c), each agent asynchronously updates and transmits the system state without waiting for its neighbors.} \label{whl2-fig:1}
\end{figure*}
Consider the MAS {consisting of} $M$ dynamically decoupled nonlinear agents. The communication topology among agents is described by a directed graph $\mathcal{G}(\mathcal{M},\mathcal{E})$ with $\mathcal{E}:=\{(i,j)\mid i,j\in \mathcal{M},i\neq j\}$ and $\mathcal{M}:=\{1,2,\dots,M\}$. Let $\mathcal{N}_i=\{j|(i,j)\in\mathcal{E}\}$ denote the set of indices of agent $i$'s neighbors and $n_i$ be the cardinality of $\mathcal{N}_i$. An edge $(i,j)\in\mathcal{E}$ means that agent $i$ can send message to agent $j$. A sequence of edges $(j_1,j_2),\dots,(j_{k-1},j_k)$ with $(j_{p-1},j_p)\in\mathcal{E}$ for all $p\in\{2,\dots,{k}\}$ is a directed path from $j_1$ to $j_k$. The graph $\mathcal{G}(\mathcal{M},\mathcal{E})$ is assumed to be connected, and each agent can broadcast the information to its neighbors. The discrete-time dynamics of agent $i$, $i\in\mathcal{M}$ takes the following form
\begin{equation}\label{whl2-eq:2}
{x}_i({t}+1)=f({x}_i(t),{u}_i(t),{d}_i(t)),\ t\in \mathbb{I},
\end{equation}
where ${x}_i\in \mathbb{R}^n$, ${u}_i\in\mathbb{R}^m$ and ${d}_i=[{w}_i,{v}_i]\in \mathbb{R}^d$ are, respectively, the state, the control input and the uncertainty. Agent $i$ is subject to the state and control input constraints
$${x}_i(t)\in \mathbb{X}_i\ \text{and} \ {u}_i(t)\in\mathbb{U}_i,$$
where the state constraint set $\mathbb{X}_i\subset \mathbb{R}^n$ is closed and the control input set $\mathbb{U}_i\subset\mathbb{R}^m$ is compact. The disturbance ${d}_i=[{w}_i,{v}_i]$ is bounded, i.e., ${d}_i\in\mathbb{D}_i\subset \mathbb{R}^d$; $\mathbb{D}_i:=\mathbb{W}_i\times\mathbb{V}_i$, with the external disturbance ${w}_i\in \mathbb{W}_i\subset \mathbb{R}^w$ and the parametric uncertainty ${v}_i\in \mathbb{V}_i\subset \mathbb{R}^v$; its upper bound is given by $\bar{d}:=\max_{{d}_i\in\mathbb{D}_i}\|{d}_i\|$. $\mathbb{W}_i$ and $\mathbb{V}_i$ are compact and contain the origin in their interiors. The function $f:\mathbb{R}^n\times\mathbb{R}^m\times\mathbb{R}^d\to\mathbb{R}^n$ is assumed to be differentiable and satisfies $f({0},{0},{0})={0}$. 

\begin{assum}\label{whl2-asm:1}
There exist constants $\nu$, $\xi\in\mathbb{R}_{\geq 0}$ such that the conditions $\|f({x},{u},{d})-f({y},{u},{d})\|\leq \nu\|{x}-{y}\|$ and $\|f({x},{u},{d})-f({x},{u},{d}')\|\leq \xi\|{d}-{d}'\|$ hold for all ${x},{y}\in \mathbb{X}_{i}$, ${u}\in\mathbb{U}_i$, ${d},{d}'\in\mathbb{D}_i$, $i\in\mathcal{M}$.
\end{assum}

\emph{Assumption \ref{whl2-asm:1}} helps to quantify an upper bound on the deviation between the newest and assumed predicted state. $\nu$ and $\xi$ can be calculated following \cite{khalil2002noninear} (see Lemmas 3.2, 3.3).

Let $t_k^i$, $k\in\mathbb{I}$ be the triggering instant of agent $i$, $i\in\mathcal{M}$. Agent $i$ measures the local system state, receives neighbors' predicted state information, and applies the calculated control action at $t_k^i$. The sampling instants can be denoted as a sequence $\{t_k^i\}$, $k\in\mathbb{I}$.

\begin{defn}\cite{xiao2008asynchronous}\label{whl2-def:4} 
The MAS in \eqref{whl2-eq:2} are synchronous if for $i,j\in\mathcal{M}$, $\{t_k^i\}=\{t_k^j\}$, i.e., $t_k^i=t_k^j$, $\forall k\in\mathbb{I}$. The MAS are said to be asynchronous if for $i,j\in\mathcal{M},i\neq j$, $\{t_k^i\}$ is independent of $\{t_k^j\}$, i.e, agents may not update their system states at the same time.
\end{defn}

The local sampling time $t_{k}^i$, $i\in\mathcal{M}$ and the communication delays $\tau_{k}^{ij}$ (from agent $j$ to $i$) satisfy the following assumption. If the sampling period and communication delays are arbitrarily large, it becomes difficult to guarantee the closed-loop stability property of the MAS without new measurements. Thus, the following assumption is given.
\begin{assum}\label{whl2-asm:2}
For agent $i$, $i\in\mathcal{M}$, the local sampling instant $t_k^i$ and the communication delays $\tau_{k}^{ij}$, $k\in\mathbb{I}$, satisfy: 1) $1\leq t_{{k}+1}^i-t_{k}^i\leq\bar{H}$; 2) $0< \tau_{k}^{ij}\leq \bar{\tau}$; 3) There is no disordering transmission among agents. Here the largest admissible sampling interval $\bar{H}\in\mathbb{I}$ and the largest communication delay $\bar{\tau}\in\mathbb{I}$ are finite.
\end{assum}

\subsection{Problem formulation}
The objective is to design a robust DMPC method for nonlinear MAS with time-varying delays, parametric uncertainties, and external disturbances, such that the MAS are robustly stabilized. Furthermore, a distributed self-triggered scheduler is developed to reduce the communication burden and the frequency of solving the corresponding optimization problems. 

\subsection{Asynchronous communication with time-varying delays}
\label{whl2-sec:III-3}
At $t_{k}^i$, agent $i$, $i\in\mathcal{M}$ {broadcasts} the newest predicted state sequence $\bm{x}_{i}^b(t_{k}^{i}):=\big({x}_i^b(\cdot|t_{k}^i)\big)$ to its neighbors as shown in Fig.~\ref{whl2-fig:1}(c), which is constructed as 
\begin{flalign} \label{whl2-eq:3}
{x}_i^b(s|t_{k}^i)=
    \begin{dcases}
        {x}_i^*(s|t_{k}^i), & s\in\mathbb{I}_{[{0},N)} \\
        {0}, & s\in\mathbb{I}_{[N,H^{i*}(t_{k}^i)+\bar{\tau}+N]},
    \end{dcases}
\end{flalign}
where ${x}_i^*(s|t_k^i)$ denotes the optimal predicted state and $H^{i*}(t_k^i)\in\mathbb{I}_{[1,\bar{H}]}$ is the optimal triggering interval at $t_k^i$. The calculation of ${x}_i^*(s|t_k^i)$ and $H^{i*}(t_k^i)$ will be introduced in Section \ref{whl2-sec:3}.

Now we describe the bounded time-varying communication delays for the asynchronous MAS. Let $\bm{x}_j^b(t_k^{ij})$ represent the newest message broadcast by agent $j,j\in \mathcal{N}_i$ at $t_k^{ij}$, where $t_{k}^{ij}:=\max\{t_{l}^j \in\mathbb{I}\mid \ t_{l}^j< t_{k}^i\}$. The communication delays can be categorized as the following two cases:\\
\textbf{Case 1}: $0\leq \tau_{k}^{ij}\leq t_{k}^i-t_{k}^{ij}$, i.e., the newest message $\bm{x}_j^b(t_{k}^{ij})$ of agent $j$ is received by agent $i$ at $t_{k}^{i}$;\\
\textbf{Case 2}: $t_{{k}}^i-t_{k}^{ij}<\tau_{k}^{ij}\leq\bar{\tau} $, i.e., the newest message $\bm{x}_j^b(t_{k}^{ij})$ cannot be received by agent $i$ at $t_{{k}}^i$, therefore, agent $i$ can only utilize the message $\bm{x}_j^b(t_{k-1}^{ij})$ broadcast at previous time instants, which includes the neighbors' state information ${x}_j^b(s|t_{k-1}^{ij}),s\in\mathbb{I}_{[0,{H}^{i*}(t^{ij}_{k-1})+\bar{\tau}+N]}$. 

Consequently, the predicted state trajectories of neighbors ${\bm{x}}_{j}(t_{k}^i),j\in\mathcal{N}_i$ used in the local optimization problem of agent $i$ in the next section can be constructed as
\begin{equation}\label{whl2-eq:4}
{\bm{x}}_{j}(t_{k}^i):=\big({x}_{j}^b(0|t_{k}^i),{x}_{j}^b(1|t_{k}^i),\cdots,{x}_{j}^b(N-1|t_{k}^i)\big).
\end{equation}

{At $t_k^i$, each agent receives its neighbors' newest predicted state sequence ${x}_j^b(\cdot|t_k^j)$, $j\in\mathcal{N}_i$ as in \eqref{whl2-eq:3} and \eqref{whl2-eq:4} to formulate the objective function for the DMPC optimization problem}
\begin{equation*}
{J}_{i,N}^{H^i}[t_{k}^i]=\sum_{s=0}^{H^i-1}\frac{1}{\hbar_i}L_i[s|t_k^i]+\sum_{s=H^i}^{N-1}L_i[s|t_k^i]+F_i[N|t_{k}^i],
\end{equation*} 
where ${J}_{i,N}^{H^i}[t_{k}^i]={J}_{i,N}^{H^i}({x}_i(t_{k}^i),{\bm{x}}_{-i}(t_{k}^{i}),\bm{u}_i(t_{k}^i),\bm{d}_i(t_{k}^i))$, $L_i[s|t_k^i]=L_i({x}_i(s|t_{k}^i),
{x}_{-i}(s|t_{k}^{i}),{u}_i(s|t_{k}^i),{d}_i(s|t_{k}^i))$, $F_i[N|t_{k}^i]=F_i({x}_i(N|t_{k}^i))$, $N$ is the prediction horizon and $H^i(t_k^i)$ denotes the triggering interval. The parameter $\hbar_i>1$ allows a trade-off between the control performance (with respect to {the optimal value of the} objective function) and the communication load (i.e., the communication frequency of each agent). ${\bm{x}}_{-i}(t_{k}^{i}):=\big(\bm{x}_{i_1}(t_{k}^{i}),\dots,\bm{x}_{i_{n_i}}(t_{k}^{i})\big)$ denote the collection of the predicted state sequence of agent $i$'s neighbors as in $(3)$, with $i_q\in\mathcal{N}_i$, $q\in\mathbb{I}_{[1,n_i]}$, ${x}_{-i}(s|t_{k}^{i}):=\big({x}_{i_1}^b(s|t_{k}^{i}),\dots,{x}_{i_{n_i}}^b(s|t_{k}^{i})\big)$, $s\in\mathbb{I}_{[0,N]}$, $\bm{u}_i(t_{k}^{i}):=\big({u}_i(0|t_k^i),\dots,{u}_i(N-1|t_k^i)\big)$ and $\bm{d}_i(t_{k}^{i}):=\big({d}_i(0|t_k^i),\dots,{d}_i(N-1|t_k^i)\big)$. The local stage cost of agent $i,i\in\mathcal{M}$ is designed as 
\begin{equation*}
\begin{aligned}
L_i[s|t_{k}^i]=&\|{{x}_i(s|t_k^i)}\|_{Q_i}^2+\|{{u}_i(s|t_k^i)}\|_{R_i}^2\\
&+\sum_{j\in\mathcal{N}_i}\|{x}_i(s|t_k^i)-{x}_j^b(s|t_k^{i})\|_{Q_{ij}}^2,
\end{aligned}
\end{equation*}
where the weighting matrices $Q_i$ and $Q_{ij}$ are symmetric and positive semidefinite, $R_i$ is symmetric and positive definite. Intuitively, the coupling cost term $\|{x}_i(s|t_k^i)-{x}_j(s|t_k^{i})\|_{Q_{ij}}^2$, $j\in\mathcal{N}_i$ {makes agents to achieve cooperation.} The local terminal cost is designed as $F_i({x}_i(N|t_k^i))=\|{{x}_i(N|t_k^i)}\|_{P_i}^2,$ where $P_i$ is symmetric and positive semidefinite.

\begin{assum}\label{whl2-asm:3}
For agent $i$, $i\in\mathcal{M}$, the local decoupled terminal set $\Omega_{i}\subseteq \mathbb{X}_{i}$ is an RPI set with a local feedback controller $\kappa_{i}({x}_i(t))\in \mathbb{U}_i$, $t\in\mathbb{I}$. There {exist a} $\mathcal{K}$ function $\sigma_i(\cdot)$ and a constant $c_4\in\mathbb{R}_{\geq 0}$ such that
$F_i({x}_i(t+1))-F_i({x}_i(t))\leq -{L}_i({x}_i(t),{x}_{-i}(t),\kappa_{i}({x}_i(t)),{d}_i(t))+\epsilon_i$, where $\epsilon_i=\sigma_i(\|{w}_i(t)\|)+c_4$, $x_{-i}(t)=0$, for all ${x}_i(t)\in{\Omega}_i$ and {${d}_i(t)\in\mathbb{D}_i$}.
\end{assum}

When ${x}_i(N|t_k^i)\in\Omega_i$, from the definition of the broadcast state trajectory ${x}_i^b(\cdot|t_{k}^i),i\in\mathcal{M}$ as in \eqref{whl2-eq:3}, it is easy to know that the state of its neighbors ${x}_j(N|t_k^i)={x}_j(N+t_k^i-t_k^j|t_k^j)={0}$, $j\in\mathcal{N}_i$. Then, the conventional design of the terminal region and terminal controller for the single system (see, e.g., \cite{liu2018robust,limon2006input}) is also applicable to the MAS. 
\section{Self-triggered Min-max DMPC}
\label{whl2-sec:3}
This section presents a solution to the problem formulated in the previous section based on the min-max DMPC. Furthermore, a distributed self-triggered scheduler for the nonlinear asynchronous MAS with bounded time-varying communication delays is designed. 

\subsection{Min-max DMPC optimization problem}
The min-max optimization problem $\mathcal{P}_i$ is designed as
\begin{subequations}\label{whl2-eq:5}
\begin{alignat}{2}
\min_{{u}_i(s|t_{k}^i)}&\{\max_{{d}_i(s|t_{k}^i)\in\mathbb{D}_i}\{\bar{J}_{i,N}^{H^i}({x}_i(t_{k}^i),{\bm{x}}_{-i}(t_{k}^{i}),{u}_i(s|t_{k}^i),{d}_i(s|t_{k}^i))\},\notag\\ 
\text{such}\ & \text{that}\ {x}_i(H^i|t_{k}^i)\in \mathbb{X}_i^{N-H^i}(\Omega_i), \forall {d}_i(s|t_{k}^i)\in \mathbb{D}_i\} \notag\\
\text{s.t.}&\ {{x}}_i(s+1|t_{k}^i)=f({x}_i(s|t_{k}^i),{u}_i(s|t_{k}^i),{d}_i(s|t_{k}^i)),\notag\\
&\  {x}_i(0|t_{k}^i)={x}_i(t_{k}^i), \label{whl2-eq:5c}\\
&\ {u}_i(s|t_{k}^i)\in \mathbb{U}_i,\label{whl2-eq:5d}\\ 
&\ {x}_i(s|t_{k}^i)\in \mathbb{X}_i,\label{whl2-eq:5e}\\
&\ \|{{x}_i(s|t_{k}^i)-{x}_i^b(s+t_{k}^{i}-t_{k-1}^{i}|t_{k-1}^{i})}\|\leq \Delta_i, \hspace{-0.2em}\label{whl2-eq:5f}
\end{alignat}
\end{subequations}
where {the triggering interval $H^i(t_k^i)$ is abbreviated as $H^i$}, $s\in\mathbb{I}_{[0,H^i)}$, $\Omega_i:=\{{x}_i\in\mathbb{X}_i\mid \|{x}_i\|_{P_i}\leq\rho_i,\ \rho_i\in\mathbb{R}_{\geq 0}\}$ is the terminal set, {$\Delta_i\in\mathbb{R}_{\geq 0}$} is a design parameter and $V_{i,N}^{H^i}({x}_i(t_{k}^i),{\bm{x}}_{-i}(t_{k}^{i}))$ is the optimal value of the objective function of $\mathcal{P}_i$. $\bar{J}_{i,N}^{H^i}({x}_i(t_{k}^i),{\bm{x}}_{-i}(t_{k}^{i}),{u}_i(s|t_{k}^i),{d}_i(s|t_{k}^i)):=\sum_{s=0}^{H^i-1}\frac{1}{\hbar_i}L_i({x}_i(t_{k}^i),{x}_{-i}(s|t_{k}^{i}),{u}_i(s|t_{k}^i),{d}_i(s|t_{k}^i))+V_{i,N-H^i}({x}_{i}(H^i|t_{k}^i),{x}_{-i}(H^i|t_{k}^{i}))$ with
\begin{equation}
\begin{aligned}
&V_{i,\ell}({x}_{i}(s|t_{k}^i),{x}_{-i}(s|t_{k}^{i}))\\
=&\min_{{\mu}_{i,\ell}(s|t_k^i)\in\mathbb{U}_i}\big\{\max_{{d}_i(s|t_k^i)\in\mathbb{D}_i}\big\{L_i({x}_i({s|t_{k}^i}),{x}_{-i}(s|t_{k}^{i}),\\
&{\mu}_{i,\ell}(s|t_{k}^i),{d}_i(s|t_{k}^i))+V_{i,\ell-1}(f({x}_i(s|t_{k}^i),{\mu}_{i,\ell}(s|t_{k}^i),\\
&{d}_i(s|t_k^i)),{x}_{-i}(s+1|t_{k}^{i}))\big\},\text{such that}\ \\
&{x}_i(s|t_{k}^i)\in \mathbb{X}_i^{\ell}(\Omega_i),\\
&f({x}_i(s|t_{k}^i),{\mu}_{i,\ell}(s|t_{k}^i),{d}_i(s|t_k^i))\subseteq \mathbb{X}_i^{\ell-1}(\Omega_i)\\
&\|{{x}_i(s|t_{k}^i)-{x}_i^b(s+t_{k}^{i}-t_{k-1}^{i}|t_{k-1}^{i})}\|\leq \Delta_i\big\},\label{whl2-eq:6}
\end{aligned}
\end{equation}
where $\mathbb{X}_i^{\ell}(\Omega_i)$ is the {$\ell$-step} robustly stabilizable set \cite{lazar2008input} of agent $i$, $\ell=N-s$; $V_{i,0}({x}_i(N|t_{k}^i))=F_i({x}_i(N|t_{k}^i))$. Solving Problem $\mathcal{P}_i$ yields the optimal control sequence $\bm{u}_{i}^*(t_{k}^i):=\big({u}^*_i(0|t_{k}^i),\dots,{u}^*_i(H^{i}-1|t_{k}^i),{\mu}_{i,\ell}^*(s|t_{k}^i),\dots,{\mu}_{i,1}^*(N-1|t_{k}^i)\big)$, where ${u}^*_i(s|t_{k}^i), s\in\mathbb{I}_{[0,H^{i})}$ is the optimal open-loop control action and ${\mu}_{i,\ell}^*(s|t_{k}^i),s\in\mathbb{I}_{[H^i,N)}$ is the optimal feedback control {law} generated by solving \eqref{whl2-eq:6} via dynamic programming. And ${d}_i^*(s|t_{k}^i),s\in\mathbb{I}_{[H^i,N)}$ is the optimal disturbance sequence.

\begin{rem}
Both the communication load and control performance are considered in the objective function of the DMPC optimization problem $\mathcal{P}_i$ when $\hbar_i>1$. In this context, the increasing value of $\hbar$ weakens the effect of the open-loop prediction (i.e., decreasing the weight of the open-loop cost function concerning the overall cost function), while guaranteeing comparable control performance compared to periodic min-max DMPC. A similar design of the parameter $\hbar_i$ was exploited for the single system in \cite{brunner2016robust,liu2018robust}. Especially, when $H^i(t_k^i)=1$, the triggered scheme becomes a standard periodic triggered scheme.
\end{rem}
\begin{rem}
The consistency constraint in \eqref{whl2-eq:5d} that specifies an upper bound of the deviation between the newest predicted states and previously broadcast states is incorporated into the local DMPC optimization problem. If the assumed state and the optimal predicted state calculated at the last time instant coincide with each other, the constraint \eqref{whl2-eq:5d} can be reduced to the compatibility constraint in \cite{dunbar2006distributed} for the synchronous MAS with delay-free and periodic communication. Hence, the proposed consistency constraint is more general and captures the compatibility constraint in \cite{dunbar2006distributed} as a special case.
\end{rem}

\subsection{Distributed self-triggered scheduler}

A distributed self-triggered scheduler is designed to generate {a sequence} $\{t_k^i\}$, based on the local system state and newest neighbors' predicted states, i.e., $t_{{k}+1}^i=t_{k}^i+H^{i*}(t_{k}^i)$ and
\begin{equation}\label{whl2-eq:7}
\begin{aligned}
{H^{i*}(t_{k}^i)}&{=\max\{H^i\in\mathbb{I}_{[1,\bar{H}]}|
V_{i,N}^{H^i}[t_{k}^i]}{\leq V_{i,N}^1[t_{k}^i]\}},
\end{aligned}
\end{equation}
where $H^{i*}(t_{k}^i)$ is the optimal triggering interval at $t_{k}^i$ and $V_{i,N}^{H^i}[t_{k}^i]=V_{i,N}^{H^i}({x}_i(t_{k}^i),{\bm{x}}_{-i}(t_{k}^{i}))$. $H^{i*}(t_{k}^i)$ is abbreviated as $H^{i*}$ hereafter. {Applying the first $H^{i*}$ open-loop control inputs $\bm{u}_i^{\text{mpc}}(t_k^i):={u}_i^{*}(s|t_{k}^i)$, $s\in\mathbb{I}_{[0,H^{i*})}$ to the system \eqref{whl2-eq:2} yields the following closed-loop system
\begin{equation}\label{whl2-eq:8}
{x}_i(t_{k}^i+s+1)=f({x}_i(t_k^i+s),{u}_i^*(s|t_k^i),{d}_i(t_k^i+s)),
\end{equation}
and ${d}_i(t_k^i+s)$, $s\in\mathbb{I}_{[0,H^{i*})}$ is the actual disturbance sequence.}
\begin{rem}
It is worth mentioning that the existing results on self-triggered DMPC in \cite{zhan2018distributed,mi2019self} are restricted to the special case of deterministic linear MAS with perfect communication. In contrast, the proposed triggered mechanism is extended to more general situations, i.e., the nonlinear MAS with two types of uncertainties and communication delays.
\end{rem}

\subsection{Self-triggered asynchronous min-max DMPC algorithm}
{The proposed self-triggered DMPC method is summarized }in \textbf{Algorithm 1}. 
\begin{algorithm}[H]
\caption{Self-triggered asynchronous min-max DMPC}\label{alg:m3pc}
\begin{algorithmic}[1]
\Require For agent $i,i\in \mathcal{M}$, the weighting matrices $Q_i$, $Q_{ij}$, $R_i$ and $P_i$; the prediction horizon $N$, the terminal set $\Omega_i$, the parameter $\Delta_i$, the initial state ${x}_i(t_{k}^i)$ and other related parameters. Set ${k}=0$, $t_k^i=0$, $H^i(t_k^i)=1$, ${x}_i^b(\cdot|t_k^i)={0}$. \hspace{-0.5em} \label{step1} 
\While{The {control action} is not stopped}
\State Sample the system state ${{x}}_i(t_{k}^i)$;
\State Receive the predicted state sequence $\bm{x}_j^b(t_{k}^{ij})$, $j\in \mathcal{N}_i$;\label{step2}
\State Construct the state sequence $\bm{x}_j(t_k^i)$ as in \eqref{whl2-eq:4};
\State Let {$H^i(t_k^i)=1$}, solve Problem $\mathcal{P}_i$ to generate ${u}_{i}^{*}(t_{k}^i)$ and $V_{i,N}^{1}$, {then set $H^i(t_k^i)=\bar{H}$}; \label{step3}
\State Solve Problem $\mathcal{P}_i$ to generate ${u}_{i}^{*}(t_{k}^i)$ and $V_{i,N}^{H^i}$;
\If{{$V_{i,N}^{H^i}> V_{i,N}^1$}}
\State {$H^i(t_k^i)=H^i(t_k^i)-1$}; {Go to Step $6$;}
\Else 
\State Generate $H^{i*}=H^i(t_k^i)$ and $V_{i,N}^{H^{i*}}$;
\EndIf   
\State {Broadcast $\bm{x}_i^b(t_{k}^i)$ as in \eqref{whl2-eq:3} to its neighbors;}
\State Apply control input $\bm{u}_i^{\text{mpc}}(t_{k}^i)$ to agent $i$;
\State {$t_{k+1}^i=t_{k}^i+H^{i*}$, ${k}={k}+1$};
\EndWhile\label{endwhile}
\end{algorithmic}
\end{algorithm}

Note that the distributed optimization problem $\mathcal{P}_i$ is solved without considering the constraint \eqref{whl2-eq:5f} in the initialization step. Besides, for each agent, the predicted states of neighbors are assumed to be zeros over the period ${[t_0^i,t_0^i+N]}$, and the communication delay at $t_0^i$ is zero, i.e., $\tau_0^{ij}=0$.
\section{Theoretical Analysis}
\label{whl2-sec:4}
\begin{lem} \label{whl2-lem:2}
Under \emph{ Assumptions~\ref{whl2-asm:1}} and \emph{\ref{whl2-asm:2}}, if $\mathcal{P}_i$ is feasible at $t_k^i$, and the condition $\Delta_i\geq\max\{\nu^{N-1-\bar{H}}\bar{\phi},\rho_i/{\underline{\lambda}(P_i^{1/2})}\}$ holds, where $\bar{\phi}:=2\sum_{s=0}^{\bar{H}+t_k^i-t_\tau^i}\nu^s\xi\bar{d}$, $t_\tau^i\in{[t_k^i-\bar{H},t_k^i]}$, $t_{k+1}^i=t_k^i+H^{i*}$, then, the inequality $\|\tilde{{x}}_i(s|t_{k+1}^i)-{x}_i^b(t_{k+1}^i+s-t_k^i|t_k^i)\|\leq\Delta_i,s\in\mathbb{I}_{[0,N]}$ holds.
\end{lem}
\begin{IEEEproof}
If $s\in\mathbb{I}_{[N-H^{i*},N]}$, it follows from \eqref{whl2-eq:3} that ${x}_i^b(t_{k+1}^i+s-t_\tau^i|t_\tau^i)=0$, we have $\|\tilde{{x}}_i(s|t_{k+1}^i)-{x}_i^b(t_{k+1}^i+s-t_\tau^i|t_\tau^i)\|=\|\tilde{{x}}_i(s|t_{k+1}^i)\|$. By the condition $\Delta_i\geq\max\{\nu^{N-1-\bar{H}}\bar{\phi},\rho_i/{\underline{\lambda}(P_i^{1/2})}\}$, it follows that $\|\tilde{{x}}_i(s|t_{k+1}^i)\|\leq\rho_i/{\underline{\lambda}(P_i^{1/2})}\leq\Delta_i$.
 
If $s=0$, by \emph{Assumption~\ref{whl2-asm:1}}, we get
\begin{equation*}
\begin{split}
&\|\tilde{{x}}_i(0|t_{k+1}^i)-{x}_i^b(t_{k+1}^i-t_\tau^i|t_\tau^i)\|\\
\leq&\nu\|{x}_i(t_{k+1}^i-1)-{x}_i^*(t_{k+1}^i-1-t_\tau^i|t_\tau^i)\|\\
&+\xi\|{d}_i(t_{k+1}^i-1)-{d}_i^*(t_{k+1}^i-1-t_\tau^i|t_\tau^i)\|\\
\leq&\nu\|{x}_i(t_{k+1}^i-1)-{x}_i^*(t_{k+1}^i-1-t_\tau^i|t_\tau^i)\|+2\xi\bar{d}.
\end{split}
\end{equation*}

Since ${x}_i(t_\tau^i)={x}_i^b(0|t_\tau^i)$, then we can obtain
\begin{equation*}
\|\tilde{{x}}_i(0|t_{k+1}^i)-{x}_i^b(t_{k+1}^i-t_\tau^i|t_\tau^i)\|\leq2\sum_{s=0}^{t_{k+1}^i-t_\tau^i}\nu^s\xi\bar{d}=\phi.
\end{equation*}

Next, we consider $s\in \mathbb{I}_{(0,N-H^{i*})}$
\begin{equation*}
\begin{split}
&\|\tilde{{x}}_i(s|t_{k+1}^i)-{x}_i^b(t_{k+1}^i+s-t_\tau^i|t_\tau^i)\|\\
=& \|f(\tilde{{x}}_i(s-1|t_{k+1}^i),{u}_i^*(t_{k+1}^i+s-1-t_\tau^i|t_\tau^i),\\
&{d}_i^*(t_{k+1}^i+s-1-t_\tau^i|t_\tau^i))\\
&-f({x}_i^*(t_{k+1}^i-1-t_\tau^i|t_\tau^i),{u}_i^*(t_{k+1}^i-1-t_\tau^i|t_\tau^i),\\
&{d}_i^*(t_{k+1}^i-1-t_\tau^i|t_\tau^i))\|\\
\leq& \nu\|\tilde{{x}}_i(s-1|t_{k+1}^i)-{x}_i^*(t_{k+1}^i+s-1-t_\tau^i|t_\tau^i)\|,
\end{split}
\end{equation*}
which implies
\begin{equation*}
\begin{split}
&\|\tilde{{x}}_i(N-H^{i*}-1|t_{k+1}^i)\\
&-{x}_i^b(t_{k+1}^i+N-H^{i*}-1-t_\tau^i|t_\tau^i)\|\\
\leq&\nu^{N-1-H^{i*}}\|\tilde{{x}}_i(0|t_{k+1}^i)-{x}_i^b(t_{k+1}^i-t_\tau^i|t_\tau^i)\|\\
\leq&\nu^{N-1-H^{i*}}\phi\\
\leq&\nu^{N-1-\bar{H}}\bar{\phi}.
\end{split}
\end{equation*}

{Then based on the condition $\Delta_i\geq\max\{\nu^{N-1-\bar{H}}\bar{\phi},$ $\rho_i/{\underline{\lambda}(P_i^{1/2})}\}$,} we have
$$\|\tilde{{x}}_i(s|t_{k+1}^i)-{x}_i^b(t_{k+1}^i+s-t_\tau^i|t_\tau^i)\|\leq\Delta_i,$$
which implies that the inequality $\|\tilde{{x}}_i(s|t_{k+1}^i)-{x}_i^b(t_{k+1}^i+s-t_k^i|t_k^i)\|\leq\Delta_i$ holds, for all $s\in\mathbb{I}_{[0,N]}$.
\end{IEEEproof}

\begin{lem} \label{whl2-lem:3}
Consider the system in \eqref{whl2-eq:2} and suppose that \emph{ Assumption~\ref{whl2-asm:3}} holds. Then, 
$V_{i,\ell+1}({x}_i(t),{x}_{-i}(t))-V_{i,\ell}({x}_i(t),{x}_{-i}(t))
\leq\max_{{d}_i}\{V_{i,\ell}({x}_i$ $(t+1),{x}_{-i}(t+1))-V_{i,\ell-1}({x}_i(t+1),{x}_{-i}(t+1))\}$, where $t\in\mathbb{I}$, $\forall{x}_i(t)\in \mathbb{X}_i^{\ell}(\Omega_i)$, $\forall{x}_j(t)\in\mathbb{X}_j^{\ell}(\Omega_j),j\in\mathcal{N}_i$, $\forall{d}_i(t)\in\mathbb{D}_i$. 
Also, 
$V_{i,\ell}({x}_i(t),{x}_{-i}(t))-V_{i,\ell-1}({x}_i(t),{x}_{-i}(t))\leq \epsilon_i$ and 
$V_{i,\ell}({x}_i(t),{x}_{-i}(t))\leq V_{i,0}({x}_i(t))+\ell\epsilon_i$.
\end{lem}
\begin{IEEEproof} 
The proof can follow Theorem 2 in \cite{limon2006input}.
\end{IEEEproof}

\begin{lem}\label{whl2-lem:4}
{For the constrained min-max DMPC optimization problem \(\mathcal{P}_i\) in \eqref{whl2-eq:5}, the inequality {\(V_{i,N}^{1}({x}_i(t_k^i),{\bm{x}}_{-i}\)} \((t_k^i))\leq V_{i,N}({x}_i(t_k^i),{\bm{x}}_{-i}(t_k^i))\) holds, where $V_{i,N}({x}_i(t_k^i),{\bm{x}}_{-i}$ $(t_k^i)):={J}_{i,N}({x}_i(t_{k}^i),{\bm{x}}_{-i}$ $(t_{k}^{i}),\bm{u}_i^*(t_{k}^i),\bm{d}_i^*(t_{k}^i))=\sum_{s=0}^{N-1}L_i$ $({x}_i^*(s|{t_{k}^i}),{x}_{-i}(s|t_{k}^{i}),{u}_i^*({s|t_{k}^i}),{d}_i^*(s|t_{k}^i))+F_i({x}_i^*(N|t_{k}^i))$.}
\end{lem}
\begin{IEEEproof}
The proof of \emph{Lemma~\ref{whl2-lem:4}} can follow the idea in \cite{liu2018robust}, so it is omitted here. 
\end{IEEEproof}
\begin{thm}\label{whl2-theorem1}
Suppose \emph{Assumptions~\ref{whl2-asm:1}}--\emph{\ref{whl2-asm:3}} hold, and Problem $\mathcal{P}_i$ is feasible at the initial time $t_0^i$. 1) Then, by application of {\bf{Algorithm 1}}, Problem $\mathcal{P}_i$ is recursively feasible at any sampling time $t_{k}^i, k\in \mathbb{I}_{[1,+\infty)}$, if the condition in \emph{Lemma \ref{whl2-lem:2}} holds. 2) Furthermore, the closed-loop uncertain system with the self-triggered min-max DMPC strategy is ISpS at triggering time instants.
\end{thm}

\begin{IEEEproof} 
\textbf{Feasibility}: For agent $i$, $i\in\mathcal{M}$, by assumption, there is a feasible solution for Problem $\mathcal{P}_i$ at $t_{k}^i$ and the recursive feasibility for all subsequent triggering instants is proven by induction. The optimal control input sequence obtained at $t_{k}^i$ is $\bm{u}_{i}^*(t_{k}^i):=\big({u}^*_i(0|t_{k}^i),\dots,{u}^*_i(H^{i*}-1|t_{k}^i),{\mu}_{i,\ell}^*(H^{i*}|t_{k}^i),\dots,{\mu}_{i,1}^*(N-1|t_{k}^i)\big)$. The first $H^{i*}$ open-loop control actions are implemented to the system in \eqref{whl2-eq:2}. At time $t_{k+1}^i$, a candidate control input sequence $\tilde{\bm{u}}_i(t_{k+1}^i)=\tilde{{u}}_i(\cdot|t_{{k}+1}^i)$ can be constructed as 
\begin{equation}\label{whl2-eq:9}
\tilde{{u}}_i(s|t_{{k}+1}^i)=
    \begin{dcases}
        {\mu}_{i,\ell}^*(H^{i*}+s|t_{k}^i), & s\in\mathbb{I}_{[0,N-H^{i*})}, \\
        \kappa_i(\tilde{{x}}_i(s|t_{k+1}^i)), & s\in\mathbb{I}_{[N-H^{i*},N)}.\\
    \end{dcases}
\end{equation}
Then the corresponding system states become $\tilde{{x}}_i(s+1|t_{k+1}^i)
={f_i(\tilde{{x}}_i(s|t_{k+1}^i),\tilde{{u}}_i(s|t_{k+1}^i),{d}_i^*(s+H^{i*}|t_k^i)),}$ where $\tilde{{x}}_i(0|t_{k+1}^i)={x}_i(t_{k+1}^i)$.
From {\emph{Assumption~\ref{whl2-asm:3}}} and the optimal feedback control input calculated at the previous time instant $t_k^i$, it follows that $\tilde{{u}}_i(s|t_{{k}+1}^i)\in \mathbb{U}_i,s\in\mathbb{I}_{[0,N)}$, the control input constraint \eqref{whl2-eq:5d} {is satisfied}. For $s\in\mathbb{I}_{[0,N-H^{i*}]}$, we have the candidate state {{$\tilde{{x}}_i(s|t_{{k}+1}^i)\in \mathbb{X}_i^{N-H^{i*}-s}(\Omega_i)\subset\mathbb{X}_i$}}; then, for $s\in\mathbb{I}_{(N-H^{i*},N]} $, under the terminal controller $\kappa_i(\tilde{{x}}_i(s|t_{k+1}^i))$ given in \emph{{Assumption~\ref{whl2-asm:3}}}, the system state always belongs to the robust invariant set $\Omega_i$. Thus, the system state constraint \eqref{whl2-eq:5e} is fulfilled. From \emph{Lemma~\ref{whl2-lem:2}}, the feasibility of the constraint \eqref{whl2-eq:5f} is guaranteed. The recursive feasibility of the proposed algorithm is established.

\textbf{Stability}: {To prove the closed-loop stability, we need to show that the optimal value function $V_{i,N}^{H^{i*}}$ is an ISpS Lyapunov function at triggering time instants. Since the stage cost function $L_i$ for agent $i,i\in \mathcal{M}$ is quadratic and {matrices $Q_i$, $Q_{ij}$ are positive definite, it can be derived that $L_i\geq \alpha_L(\|x_i(t_{k}^i)\|)$, where $\alpha_L(\|x_i(t_{k}^i)\|)=\underline{\lambda}(Q_i)\|x_i(t_{k}^i)\|^2$ is a $\mathcal{K}_{\infty}$ function}. Then we obtain $V_{i,N}^{H^{i*}}\geq\alpha_L(\|x_i(t_{k}^i)\|)$.} The similar way to establish the upper bound of $V_{i,N}^{H^{i*}}$ in \cite{limon2006input} is adopted here. Define a set $B_{i,r}=\{{x}_i\in \mathbb{R}^n\mid \|{{x}_i}\|\leq r_i\}\subseteq \Omega_i$. Due to the compactness of $\mathbb{X}_i$ and $\mathbb{U}_i$, the optimal value of the min-max DMPC cost function is upper bounded, i.e., $V_{i,N}^{H^{i*}}({x}_i(t_{k}^i),{\bm{x}}_{-i}(t_{k}^i))\leq \bar{V}_{i,N}$. If ${x}_i(t_{k}^i)\in\Omega_i$, by {\emph{Assumption~\ref{whl2-asm:3}}} and \emph{Lemma \ref{whl2-lem:3}}, one can have
\begin{equation}\label{whl2-eq:10}
\begin{aligned}
&F_i({x}_i(s+1|t_{k}^i))-F_i({x}_i(s|t_{k}^i))\\
\leq& -L_i({x}_i(s|t_{k}^i),{x}_{-i}(s|t_{k}^i),\kappa_i({x}_i(s|t_{k}^i)),{d}_i(s|t_k^i))+\epsilon_i. 
\end{aligned}
\end{equation} 

By summing up \eqref{whl2-eq:10} from $s=0$ to $N$, we get
\begin{equation*}
\begin{aligned}
&\sum_{s=0}^{N-1}L_i({x}_i(s|t_{k}^i),{x}_{-i}(s|t_{k}^i),\kappa_i({x}_i(s|t_{k}^i)),{d}_i(s|t_k^i))\\
&+F_i({x}_i(N|t_{k}^i))\\
\leq&F_i({x}_i(t_{k}^i))+N\epsilon_i,
\end{aligned}
\end{equation*}
where $F_i({x}_i(t_{k}^i))=F_i({x}_i(0|t_{k}^i))$.

If the distributed triggering condition \eqref{whl2-eq:7} is satisfied, i.e., $V_{i,N}^{H^{i*}}({x}_i(t_{k}^i),{\bm{x}}_{-i}(t_{k}^i))\leq V_{i,N}^1({x}_i(t_{k}^i),{\bm{x}}_{-i}(t_{k}^i))$. {In the view of the definition of the terminal cost function, we have $F_i({x}_i(t_k^i))\leq \bar{\alpha}_F(\|{x}_i(t_k^i)\|)$, where $\bar{\alpha}_F(\|x_i(t_{k}^i)\|)=\overline{\lambda}(P_i)\|x_i(t_{k}^i)\|^2$ is a $\mathcal{K}_\infty$ function.} Then,
\begin{equation}\label{whl2-eq:11}
\begin{aligned}
V_{i,N}^{H^{i*}}({x}_i(t_{k}^i),{\bm{x}}_{-i}(t_{k}^i))\leq& V_{i,N}^1({x}_i(t_{k}^i),{\bm{x}}_{-i}(t_{k}^i))\\
\leq& V_{i,N}({x}_i(t_{k}^i),{\bm{x}}_{-i}(t_{k}^i))\\
\leq& \bar{\alpha}_F(\|{x}_{i}(t_{k}^i)\|)+N\epsilon_i.
\end{aligned}
\end{equation} 
The second inequality follows from the fact in \emph{Lemma \ref{whl2-lem:4}}. If ${x}_i(t_{k}^i)\in\mathbb{X}_i^{N}(\Omega_i)\backslash \Omega_i$, it implies that $\bar{\alpha}_F(\|{{x}_{i}(t_{k}^i)}\|)\geq \bar{\alpha}_F(r_i)$, where $\mathbb{X}_i^{N}(\Omega_i)$ is the $N$-step robustly stabilizable set of agent $i$. And thus
\begin{equation}\label{whl2-eq:12}
\begin{aligned}
V_{i,N}^{H^{i*}}({x}_i(t_{k}^i),{\bm{x}}_{-i}(t_{k}^i))\leq& \bar{V}_{i,N}\frac{\bar{\alpha}_F(\|{{x}_{i}}(t_{k}^i)\|)}{\bar{\alpha}_F(r_i)}\\
\leq&\theta_i\bar{\alpha}_F(\|{{x}_{i}}(t_{k}^i)\|)+N\epsilon_i,
\end{aligned}
\end{equation}
where $\theta_i=\max\{1,\frac{\bar{V}_{i,N}}{\bar{\alpha}_F(r_i)}\}$.

At time $t_k^i$, the {sequence of optimal control policies} $\bm{u}_i^*(t_k^i)$ for Problem $\mathcal{P}_i$ can steer ${x}_i(t_k^i)$ of agent $i,i\in \mathcal{M}$ to the terminal set $\Omega_i$ in $N$ steps {under} the disturbance sequence $\bm{d}_i^*(t_k^i)=\big({d}_i^*(0|t_k^i),\dots,{d}_i^*(N-1|t_k^i)\big)$. If the control actions $\bm{u}_i^{\text{mpc}}(t_{k}^i)$ and the actual disturbance $\bm{d}_i^a(t_{k}^i)$ are applied, then agent $i$ evolves to ${x}_i(t_{{k}+1}^i)$, where $\bm{d}_i^a(t_{k}^i)=\big({d}_i(t_k^i),{d}_i(t_k^i+1),\dots,{d}_i(t_k^i+H^{i*}-1)\big)$. For agent $j$, $j\in\mathcal{N}_i$, the broadcast state evolves to ${x}_j^*(H^{i*}|t_{{k}}^i)$ with the control actions ${u}_j^{*}(s|t_{k}^i)$ and the disturbance input ${d}_j^*(s|t_{k}^i)$, $s\in\mathbb{I}_{[0,H^{i*})}$. 

Then, it is easy to have 
\begin{equation*}\label{whl2-eq:15}
\begin{aligned}
&{J}_{i,N-H^{i*}}({x}_i(t_{k+1}^i),\bm{x}_{-i}(t_{k+1}^i),\bm{\mu}_{i}^*(t_k^i),\bm{d}_i'(t_k^i))\\
= &{J}_{i,N}^{H^{i*}}({x}_i(t_{k}^i),\bm{x}_{-i}(t_{k}^i),\bm{u}_{i}^*(t_k^i),\bm{d}''_i(t_k^i))\\
& -\frac{1}{\hbar_i}\sum_{s=0}^{H^{i*}-1}L_i({x}_i(s|t_{k}^i),{x}_{-i}(s|t_{k}^i),{u}_i^*(s|t_{k}^i),\bm{d}_i^a(t_k^i)),
\end{aligned}
\end{equation*}
where $\bm{\mu}_{i}^*(t_k^i)=\big({\mu}_{i,N-H^{i*}}^*(H^{i*}|t_k^i),\dots,{\mu}_{i,1}^*(N-1|t_k^i)\big)$, $\bm{d}_i'(t_k^i)=\big({d}_i^*(0|t_{k+1}^i),\dots,{d}_i^*(N-H^{i*}-1|t_{k+1}^i)\big)$ and $\bm{d}_i''(t_k)=\big({d}_i(t_k^i),\dots,{d}_i(t_k^i+H^{i*}-1),{d}_i^*(0|t_{k+1}^i),\dots,{d}_i^*(N-H^{i*}-1|t_{k+1}^i)\big)$. Then, from the above equality, we get
\begin{equation}\label{whl2-eq:13}
\begin{aligned}
&{J}_{i,N-H^{i*}}({x}_i(t_{k+1}^i),\bm{x}_{-i}(t_{k+1}^i),\bm{\mu}_{i}^*(t_k^i),\bm{d}_i'(t_k^i))\\
\leq&{J}_{i,N}^{H^{i*}}({x}_i(t_{k}^i),\bm{x}_{-i}(t_{k}^i),\bm{u}_{i}^*(t_k^i),\bm{d}''_i(t_k^i))\\
&-H^{i*}/\hbar_i\alpha_L(\|{x}_i(t_{k}^i)\|),
\end{aligned}
\end{equation}
Further, the system state ${x}_i(t_{k+1}^i)$ can be steered into the terminal set $\Omega_i$ in $N-H^{i*}$ steps by the candidate control input $\tilde{\bm{u}}_i(t_{k+1}^i)$. Then, by the inequality \eqref{whl2-eq:13}, and \emph{Assumption~\ref{whl2-asm:3}},
\begin{equation}\label{whl2-eq:14}
\begin{aligned}
&{{J}_{i,N}}({x}_i(t_{k+1}^i),\bm{x}_{-i}(t_{k+1}^i),\tilde{\bm{u}}_{i}(t_{k+1}^i),\bm{d}_i^*(t_{k+1}^i))\\
\leq&{J}_{i,N}^{H^{i*}}({x}_i(t_{k}^i),\bm{x}_{-i}(t_{k}^i),\bm{u}_{i}^*(t_k^i),\bm{d}_i''(t_k^i))\\
&-H^{i*}/\hbar_i\alpha_L(\|{x}_i(t_{k}^i)\|)+H^{i*}\epsilon_i,
\end{aligned}
\end{equation} 
where $\bm{d}_i^*(t_{k+1}^i)=\big({d}_i^*(0|t_{k+1}^i),\dots,{d}_i^*(N-1|t_{k+1}^i)\big)$. 

Consider the time-varying communication delays in {\bf{Case 2}}, i.e., $t_{{k+1}}^i-t_{k+1}^{ij}<\tau_{k+1}^{ij}\leq\bar{\tau} $, the predicted states of neighbors transmitted at the previous triggering time instant $\bm{x}_{-i}(t_{k+1}^i)=\big({x}_{-i}(H^{i*}|t_k^i),\dots,{x}_{-i}(H^{i*}+N-1|t_k^i)\big)$ will be used. Based on the triggering condition and \emph{Lemma~\ref{whl2-lem:4}}, we obtain
\begin{equation}\label{whl2-eq:15}
\begin{aligned}
&V_{i,N}^{H^{i*}(t^i_{{k}+1})}({x}_i(t_{{k}+1}^i),{\bm{x}}_{-i}(t_{{k}+1}^i))\\
\overset{\eqref{whl2-eq:7}}{\leq}& V_{i,N}^{1}({x}_i(t_{{k}+1}^i),{\bm{x}}_{-i}(t_{{k}+1}^i))\\
\leq& V_{i,N}({x}_i(t_{{k}+1}^i),{\bm{x}}_{-i}(t_{{k}+1}^i))\\
=& J_{i,N}({x}_i(t_{{k}+1}^i),{\bm{x}}_{-i}(t_{{k}+1}^i),\bm{u}_i^*(t_{k+1}^i),\bm{d}_i^*(t_{k+1}^i))\\
\leq&{J}_{i,N}({x}_i(t_{k+1}^i),\bm{x}_{-i}(t_{k+1}^i),\tilde{\bm{u}}_{i}(t_{k+1}^i),\bm{d}_i^*(t_{k+1}^i))\\
\overset{\eqref{whl2-eq:14}}{\leq}&{J}_{i,N}^{H^{i*}}({x}_i(t_{k}^i),\bm{x}_{-i}(t_{k}^i),\bm{u}_{i}^*(t_k^i),\bm{d}_i''(t_k^i))\\
&-H^{i*}/\hbar_i\alpha_L(\|{x}_i(t_{k}^i)\|)+H^{i*}\epsilon_i\\
\leq&V_{i,N}^{H^{i*}}({x}_i(t_{k}^i),{\bm{x}}_{-i}(t_{k}^i))-H^{i*}/\hbar_i\alpha_L(\|{x}_i(t_{k}^i)\|)\\
&+H^{i*}\epsilon_i.
\end{aligned}
\end{equation}
The last inequality is derived from the fact that $\bm{d}_i''(t_k^i)$ is not the optimal solution of Problem $\mathcal{P}_i$.

Consider the time-varying communication delays in {\bf{Case 1}}, i.e., $\tau_{k+1}^{ij}\leq t_{{k+1}}^i-t_{k+1}^{ij}$, the newest predicted states of neighbors $\bm{x}'_{-i}(t_{k+1}^i)=\big({x}_{-i}(0|t_{k+1}^i),\dots,{x}_{-i}(N-1|t_{k+1}^i)\big)$ will be received and used by agent $i$. Because of the triangle inequality and the consistency constraint \eqref{whl2-eq:5f}, we have
\begin{equation}\label{whl2-eq:16}
\begin{aligned}
&\|{{x}_i(s|t_{{k}+1}^i)-{x}_{j}(s|t_{{k}+1}^i)}\|\\
\leq&\|{{x}_i(s|t_{{k}+1}^i)-{x}_j^b(H^{i*}+s|t_{k}^i)}\|\\
&+\|{{x}_j(s|t_{{k}+1}^i)-{x}_j^b(H^{i*}+s|t_{k}^i)}\|\\
\overset{\eqref{whl2-eq:5f}}{\leq}&\|{{x}_i(s|t_{{k}+1}^i)-{x}_j^b(H^{i*}+s|t_{k}^i)}\|+\Delta_j.
\end{aligned}
\end{equation}
Since $x_i(\cdot|t_k^i)\in\mathbb{X}_i$, there exists a constant $\Lambda_i>0$ such that $\|{x}_i(s|t_k^i)\|\leq\Lambda_i$, $k\in\mathbb{I}$. Then it can be obtained that
\begin{equation}\label{whl2-eq:17}
\begin{aligned}
&\|{{x}_i(s|t_{{k}+1}^i)-{x}_{j}(s|t_{{k}+1}^i)}\|^2_{Q_{ij}}\\
\leq&\|{{x}_i(s|t_{{k}+1}^i)-{x}_j^b(H^{i*}+s|t_{k}^i)}\|^2_{Q_{ij}}\\
&+2\bar{\lambda}(Q_{ij})\|{{x}_i(s|t_{{k}+1}^i)-{x}_j^b(H^{i*}+s|t_{k}^i)}\|\\
&\cdot\|{{x}_j(s|t_{{k}+1}^i)-{x}_j^b(H^{i*}+s|t_{k}^i)}\|\\
&+\|{{x}_j(s|t_{{k}+1}^i)-{x}_j^b(H^{i*}+s|t_{k}^i)}\|^2_{Q_{ij}}\\
\overset{\eqref{whl2-eq:5f}}{\leq}&\|{{x}_i(s|t_{{k}+1}^i)-{x}_j^b(H^{i*}+s|t_{k}^i)}\|^2_{Q_{ij}}\\
&+4\bar{\lambda}(Q_{ij})\Lambda_i\Delta_j+\bar{\lambda}(Q_{ij})\Delta_j^2.
\end{aligned}
\end{equation}

From the triggering condition and \emph{Lemma~\ref{whl2-lem:3}}, we obtain that
\begin{equation*}
\begin{aligned}
&V_{i,N}^{H^{i*}(t^i_{{k}+1})}({x}_i(t_{{k}+1}^i),{\bm{x}}'_{-i}(t_{{k}+1}^i))\\
\overset{\eqref{whl2-eq:7}}{\leq}& V_{i,N}^{1}({x}_i(t_{{k}+1}^i),{\bm{x}}'_{-i}(t_{{k}+1}^i))\\
\leq& V_{i,N}({x}_i(t_{{k}+1}^i),{\bm{x}}'_{-i}(t_{{k}+1}^i))\\
=& J_{i,N}({x}_i(t_{{k}+1}^i),{\bm{x}}_{-i}'(t_{{k}+1}^i),\bm{u}_i^*(t_{k+1}^i),\bm{d}_i^*(t_{k+1}^i))\\
\leq&{J}_{i,N}({x}_i(t_{k+1}^i),\bm{x}'_{-i}(t_{k+1}^i),\tilde{\bm{u}}_{i}(t_{k+1}^i),\bm{d}_i^*(t_{k+1}^i))\\
\overset{\eqref{whl2-eq:17}}{\leq}&{J}_{i,N}({x}_i(t_{k+1}^i),\bm{x}_{-i}(t_{k+1}^i),\tilde{\bm{u}}_{i}(t_{k+1}^i),\bm{d}_i^*(t_{k+1}^i))\\
&+(N-H^{i*})\bar{\lambda}(Q_{ij})\sum_{j\in\mathcal{N}_i}(4\Lambda_i\Delta_j+\Delta_j^2)\\
\overset{\eqref{whl2-eq:14}}{\leq}&{J}_{i,N}^{H^{i*}}({x}_i(t_{k}^i),\bm{x}_{-i}(t_{k}^i),\bm{u}_{i}^*(t_k^i),\bm{d}_i''(t_k^i))\\
&-H^{i*}/\hbar_i\alpha_L(\|{x}_i(t_{k}^i)\|)+\Xi_i\\
\leq&V_{i,N}^{H^{i*}}({x}_i(t_{k}^i),{\bm{x}}_{-i}(t_{k}^i))-H^{i*}/\hbar_i\alpha_L(\|{x}_i(t_{k}^i)\|)+\Xi_i,\\
\end{aligned}
\end{equation*}
where $\Xi_i=(N-H^{i*})\bar{\lambda}(Q_{ij})\sum_{j\in\mathcal{N}_i}(4\Lambda_i\Delta_j+\Delta_j^2)+H^{i*}\epsilon_i$. By now, we have shown that $V_{i,N}^{H^{i*}}({x}_i(t_{k}^i),$ ${\bm{x}}_{-i}(t_{k}^i))$ is an ISpS Lyapunov function. Based on \emph{Lemma \ref{whl2-lem:1}}, it can be concluded that the closed-loop system in \eqref{whl2-eq:8} is ISpS with respect to $\Omega_i$ at triggering instants. This concludes the proof. 
\end{IEEEproof}
\section{Numerical Example}
\label{whl2-sec:5}
Three group of simulations are conducted i.e., periodic min-max DMPC (P-DMPC), self-triggered min-max DMPC without delays (ST-DMPC) and self-triggered min-max DMPC with delays (ST-DMPC-D). Consider MAS with five agents adopted in \cite{liu2018robust}, and agent $i$, $i\in\mathcal{M}$ is characterized by
\begin{equation*} 
\begin{split}
{x}_{i,1}^+=&x_{i,1}+ T(x_{i,2}),\\
{x}_{i,2}^+=&x_{i,2}-\frac{T}{m_i}\big (k'_ie^{-x_{i,1}}x_{i,1}+{h'_i}x_{i,2}-u_i+v_ix_{i,2}-w_i\big ),
\end{split}
\end{equation*}
in which $m_i=1 \text{kg}$, $k'_i=0.33 \text{N/m}$, $h'_i=1.1 \text{Ns/m}$ and $T=0.3\text{s}$. The uncertainties are bounded by $-0.1\leq v_i\leq 0.1$, $-0.15\leq w_i\leq 0.15$. The control input and state constraints are given by $-4\text{N}\leq u_i\leq4\text{N}$, $-1.95\text{m}\leq x_{i,1}\leq 1.95\text{m}$. The Lipschitz constants can be calculated as $\nu=1.23$ and $\xi=0.42$ (see Lemma 3.2 in \cite{khalil2002noninear}). The delay $\tau\in\mathbb{I}_{[1,\bar{\tau}]}$ is randomly generated, with $\bar{\tau}=3$. Other parameters are designed as follows: ${Q}_i=\text{diag}(0.6,0.6)$, ${Q}_{ij}=\text{diag}(0.5,0.5)$, $P_i=[8.05, 2.90;2.90, 3.48]$, ${R}_i=1$, $N=5$, $\bar{H}=4$, $\hbar_i=1.1$, ${\Delta}_i=3.58$ and $\rho_i=\sqrt{6.0}$. The terminal control law is $\kappa_i({x}_i)=[-0.87, -1.04]{x}_i$. The feedback control policy $\mu_i({x}_i)=a\kappa_i({x}_i)+b\|{{x}_i}\|^2+c$, where $a,b,c\in \mathbb{R}$ are decision variables. The parametric uncertainties $v_i(t)$ and external disturbances $w_i(t)$ are $0.1\sin({t}/{4\pi})$ and $0.15\cos({t}/{3\pi})$, respectively. $\mathcal{G}$ is described by $\mathcal{N}_1=\{2\}$, $\mathcal{N}_2=\{1,5\}$, $\mathcal{N}_3=\{2,4\}$, $\mathcal{N}_4=\{3\}$, $\mathcal{N}_5=\{2\}$. The initial states are given as follows: ${x}_1= [1.5, 0.7]^\text{T}$, ${x}_2 = [-0.5, -1.1]^\text{T}$, ${x}_3=[-2.0, 0.5]^\text{T}$, ${x}_4 = [0.7, -1.0]^\text{T}$, ${x}_5=[1.95, 0]^\text{T}$.

\graphicspath{{./Fig/}} 
\begin{figure}[!ht]
\includegraphics[width=0.7\columnwidth]{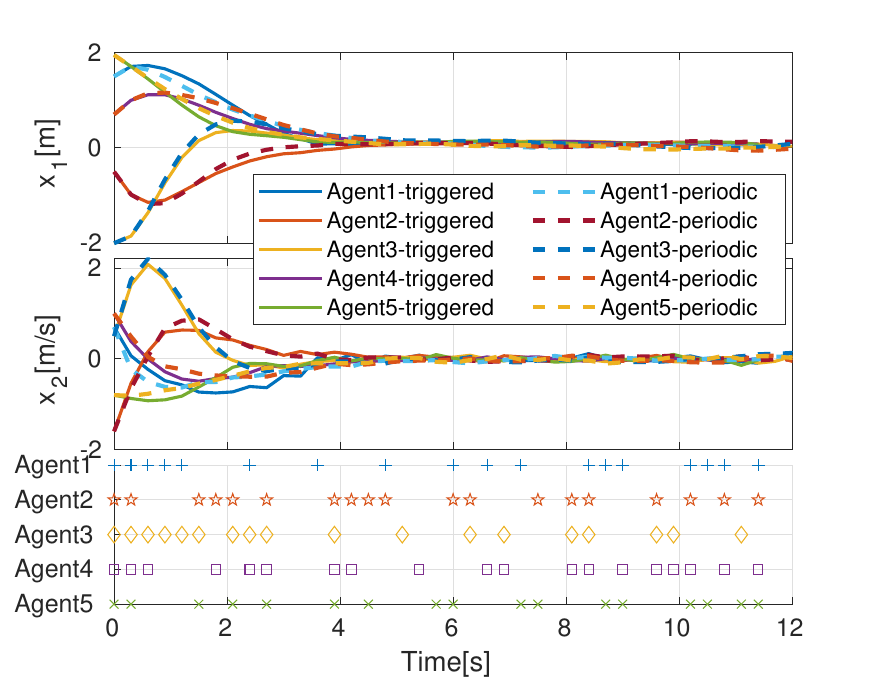} 
\centering
\caption{State trajectories and triggering instants of P-DMPC and ST-DMPC. \textbf{Top:} The displacements. \textbf{Middle:} The velocities. \textbf{Bottom:} The triggering instants.}
\label{whl2-fig:3}
\end{figure}

\begin{figure}[!ht]
\includegraphics[width=0.7\columnwidth]{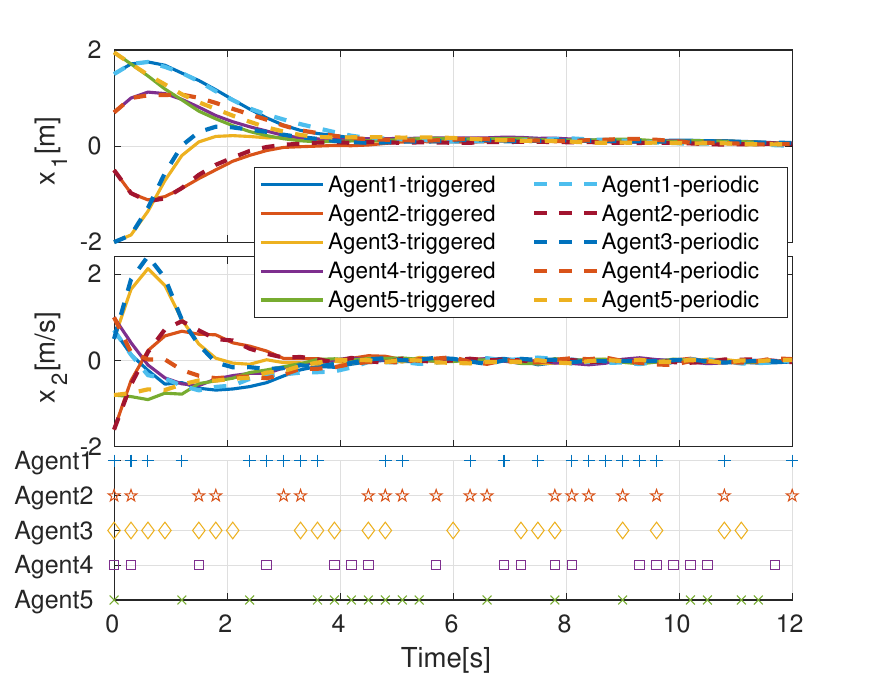}
\centering
\caption{State trajectories and triggering instants of P-DMPC and ST-DMPC-D. \textbf{Top:} The displacements. \textbf{Middle:} The velocities. \textbf{Bottom:} The triggering instants.}
\label{whl2-fig:5}
\end{figure}

Fig.~\ref{whl2-fig:3} depicts five agents' state trajectories and triggering instants without communication delays under the P-DMPC method and ST-DMPC.  The state trajectories of five agents with time-varying communication delays are shown in Fig.~\ref{whl2-fig:5}. 

{Let $\bar{T}:=\sum_{i\in\mathcal{M}}{T_\text{sim}}/{Ms_i}$ denote the average sampling time, where $s_i$ is total triggering times of agent $i$ during the simulation time $T_\text{sim}$. The control performance index is defined by $\bar{J}:={\sum_{i\in\mathcal{M}}J_i}/{M}$, $J_i=\sum_{t=0}^{T_\text{sim}}\|{x}_i(t)\|^2_{Q_i}+\sum_{j\in\mathcal{N}_i}\|{x}_i(t)-{x}_j(t)\|^2_{Q_{ij}}+\|{u}_i(t)\|^2_{R_i}$.} From Table~1, it can be seen that the communication load of MAS is significantly reduced by using ST-DMPC(-D) while achieving comparable control performance of P-DMPC. 

\graphicspath{{./Fig/}}

\begin{table}[!ht]
\caption{Performance comparison.}
\begin{tabular}{l c c}
\toprule
Method  & { Average sampling time}& Control performance index \\
\midrule
P-DMPC & {0.3000} & 51.3016\\
ST-DMPC   & {0.6303} &52.2235 \\
ST-DMPC-D   &{0.6105} & {51.4515} \\
\bottomrule
\end{tabular}
\label{whl2-tab:2}
\end{table}
\section{Conclusion}
\label{whl2-sec:6}
This paper proposed a self-triggered min-max DMPC for nonlinear perturbed MAS with communication delays. A novel consistency constraint was designed to force the deviation between the newest and previously predicted states to lie in a prescribed region. The proposed method significantly reduced the communication load with slightly affected closed-loop performance than the periodic DMPC method. As mentioned above, the control inputs were obtained by solving the min-max optimal control problem at triggering instants, which, however, was the main limitation of the proposed method due to the lack of efficient solvers to the nonlinear min-max optimization problem. Future work will focus on reducing the computational burden of the min-max DMPC.
\bibliographystyle{IEEEtran}
\bibliography{IEEEabrv,whl2_trigger}   

\begin{IEEEbiography}[{\includegraphics[width=1.1in,height=1.25in,clip,keepaspectratio]{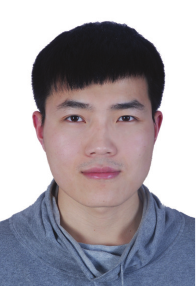}}]
{Henglai Wei} (S'17) received his B.S. (Hons.) and M.S. degrees in control theory from Northwestern Polytechnical University, Xi'an, China, in 2014 and 2017, respectively. He is currently working toward his Ph.D. degree in control theory with the University of Victoria, Canada. His current research interests include distributed model predictive control and optimization with applications on autonomous marine robots. 
\end{IEEEbiography}
\begin{IEEEbiography}[{\includegraphics[width=1in,height=1.25in,clip,keepaspectratio]{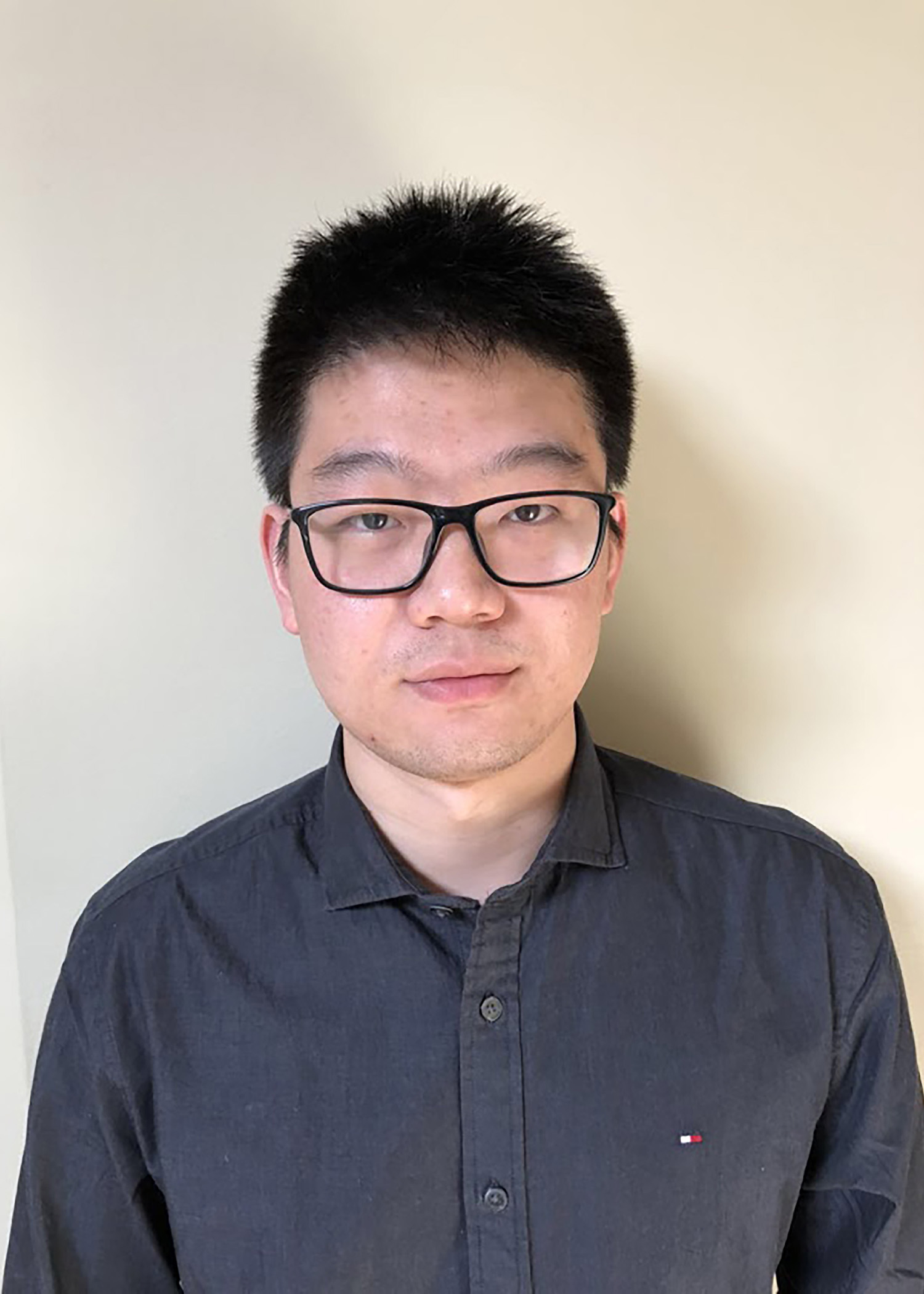}}]
{Kunwu Zhang} (S'16) received the B.Eng. degree in electrical engineering and automation from Hubei University of Science and Technology, Xianning, China, in 2013, and the M.A.Sc. degree in mechanical engineering from the University of Victoria, Victoria, BC, Canada, in 2016, where he is currently working toward the Ph.D. degree in the Department of Mechanical Engineering. 
His current research interests include adaptive control, model predictive control, optimization and robotic systems.
\end{IEEEbiography}
\begin{IEEEbiography}[{\includegraphics[width=1in,height=1.25in,clip,keepaspectratio]{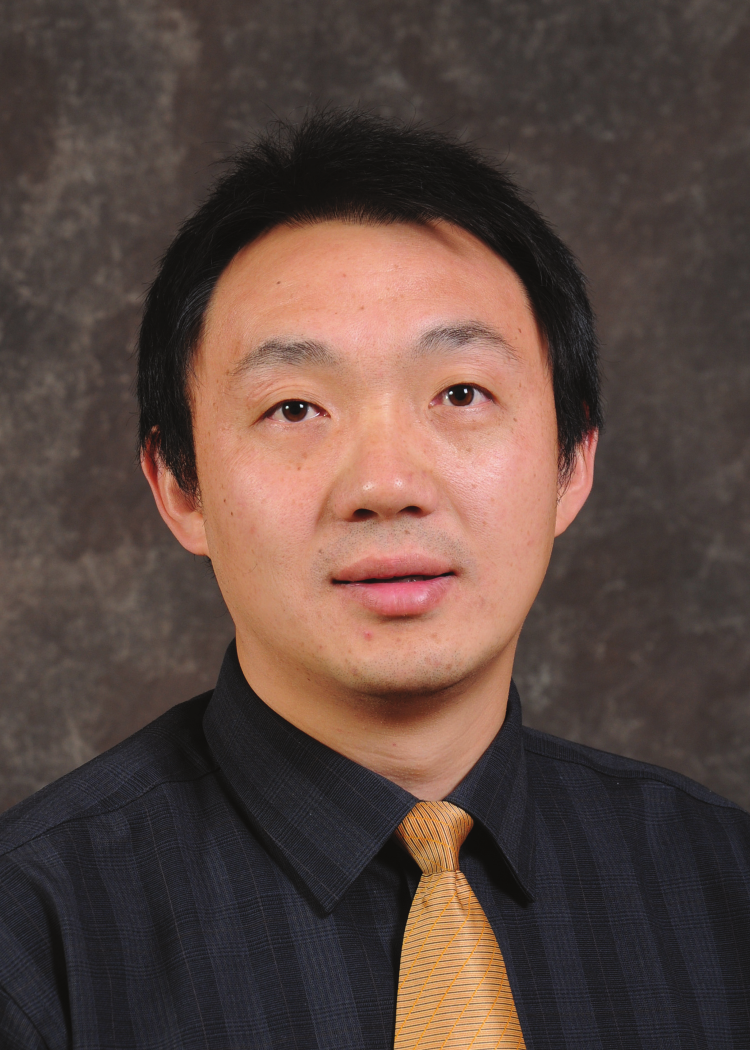}}]
{Yang Shi} (SM’09-F’17) received the Ph.D. degree in electrical and computer engineering from the University of Alberta, Edmonton, AB, Canada, in 2005. From 2005 to 2009, he was an Assistant Professor and Associate Professor in the Department of Mechanical Engineering, University of Saskatchewan, Saskatoon, SK, Canada. In 2009, he joined the University of Victoria, and now he is a Professor in the Department of Mechanical Engineering, University of Victoria, Victoria, BC, Canada. His current research interests include networked and distributed systems, model predictive control (MPC), cyber-physical systems (CPS), robotics and mechatronics, control of autonomous systems (AUV and UAV), and energy system applications.
Dr. Shi received the University of Saskatchewan Student Union Teaching Excellence Award in 2007, and the Faculty of Engineering Teaching Excellence Award in 2012 at the University of Victoria (UVic). He is the recipient of the JSPS Invitation Fellowship (short-term) in 2013, the UVic Craigdarroch Silver Medal for Excellence in Research in 2015, the 2017 IEEE Transactions on Fuzzy Systems Outstanding Paper Award, the Humboldt Research Fellowship for Experienced Researchers in 2018. Currently he serves as a member of the IEEE IES Administrative Committee, the Chair of IEEE IES Technical Committee on Industrial Cyber-Physical Systems, and Co-Editor-in-Chief for IEEE Transactions on Industrial Electronics.  He also serves as Associate Editor for Automatica, IEEE Transactions on Control Systems Technology, IEEE/ASME Transactions on Mechatronics, IEEE Transactions on Cybernetics, etc. He is General Chair of the 2019 International Symposium on Industrial Electronics (ISIE) and the 2021 International Conference on Industrial Cyber-Physical Systems (ICPS). 
He is a Fellow of IEEE, ASME, CSME, and Engineering Institute of Canada (EIC), and a registered Professional Engineer in British Columbia, Canada.
\end{IEEEbiography}
\end{document}